\documentclass{aa}  

\usepackage{graphicx}
\usepackage{txfonts}
\usepackage{hyperref}
\usepackage[normalem]{ulem}

\begin{document}

   \title{ISOSCELES project: A grid-based quantitative spectroscopic analysis of massive stars}

   \author{I. Araya
          \inst{1}
          \and
          M. Curé
          \inst{2}
          \and
          N. Machuca          
          \inst{2}
          \and
          R. O. J. Venero                
          \inst{3,4}\thanks{Member of the Carrera del Investigador Cient\'{\i}fico, CONICET, Argentina}      
          \and
          S. Cuéllar
          \inst{1}       
          \and          
          C. Arcos
          \inst{2}
          \and
          L. S. Cidale 
          \inst{3,4}$^\star$       
          }

   \institute{Centro Multidisciplinario de F\'isica, Vicerrector\'ia de Investigaci\'on, Universidad Mayor, 8580745 Santiago, Chile\\
              \email{ignacio.araya@umayor.cl}
         \and
             Instituto de F\'{\i}sica y Astronom\'{\i}a, Facultad de Ciencias, Universidad de Valpara\'{\i}so,
             Av. Gran Breta\~na 1111, Valpara\'{\i}so, Chile
         \and
         Departamento de Espectroscop\'{\i}a, Facultad de Ciencias Astron\'omicas y Geof\'{\i}sicas, Universidad Nacional de La Plata, Paseo del Bosque S/N, BF1900FWA La Plata, Buenos Aires, Argentina
         \and
          Instituto de Astrof\'{\i}sica de La Plata, CCT La Plata, CONICET-UNLP, Paseo del Bosque S/N, BF1900FWA La Plata, Buenos Aires, Argentina
             }

   \date{Received ; accepted }

  \abstract
   {Massive stars play a fundamental role in galactic evolution through their strong stellar winds, chemical enrichment, and feedback mechanisms. Accurate modelling of their atmospheres and winds is critical for understanding their physical properties and evolutionary pathways. Traditional spectroscopic analyses often rely on the $\beta$-law approximation for wind-velocity profiles, which may not capture the complexity of observed phenomena.}
   {This study aims to introduce and validate the grId of Stellar atmOSphere and hydrodynamiC modELs for massivE Stars (ISOSCELES), a grid-based framework for the quantitative spectroscopic analysis of massive stars. The project leverages hydrodynamic wind solutions derived from the m-CAK theory, including both fast and $\delta$-slow solutions, to improve the accuracy of derived stellar and wind parameters.}
   {We constructed a comprehensive grid of models based on hydrodynamic wind solutions from the {\sc Hydwind} code and synthetic spectral line profiles generated by the {\sc Fastwind} code. The grid spans a broad parameter space covering OBA-type stars with solar metallicity. A semi-automatic fitting procedure was developed to analyse key spectral lines and derive the stellar and wind parameters.}
   {Applying ISOSCELES to six stars demonstrates its ability to reproduce observed spectral profiles with high fidelity. The $\delta$-slow solution proved effective for two early-type B supergiants. The grid also highlights the difference of using the $\beta$-law in modelling stellar winds compared with the m-CAK wind solutions.}
   {The ISOSCELES database represents a step forward in quantitatively analysing massive stars, offering an alternative to the $\beta$-law approximation. Future work will address the inclusion of UV lines and metallicity effects to further refine its applicability across diverse stellar populations.}

    \authorrunning{Araya et. al.}

   \maketitle
%
\nolinenumbers
\section{Introduction}
  Due to their extreme properties and role in galactic evolution, massive stars are of fundamental importance to our understanding of the Universe. They are the primary source of chemical enrichment of the interstellar medium, which is released during stellar outflows and supernova explosions. The individual evolution of a massive star is mainly controlled by its stellar wind combined with stellar rotation. Winds from massive stars are called line-driven winds; they are driven by intense radiation pressure where the radiation field transfers momentum to the stellar plasma by scattering in metal lines \citep{lucy1970}. Currently, the hydrodynamic theory that describes these winds (m-CAK theory) is based on the pioneering work of \citet{castor1974}, \citet{castor1975}, and \citet{abbott1982} and the improvements made by \citet{friend1986} and \citet{ppk1986}; see a review from \citet{cure2023}. From the standard m-CAK theory, the line-force parameters ($\alpha$, $k$, and $\delta$) provide scaling laws for the mass-loss rate ($\dot{M}$) and terminal velocity ($v_\infty$) of the wind \citep{kudritzki1989}. The parameters $\alpha$ and $k$ describe the statistical dependence of the number of lines on frequency position and line strength \citep{puls2000}, while the $\delta$ parameter is related to changes in the ionisation throughout the wind. \citet{cure2011} found, in the frame of this m-CAK theory, a new spherical hydrodynamic solution (called the $\delta$-slow solution), which is different from the standard m-CAK solution (called the fast solution).
  This solution is obtained when high values of the $\delta$  parameter ($\gtrsim 0.3$) are considered. Although this value might seem high, \citet{puls2000} showed that for winds with neutral hydrogen as a trace element, the $\delta$ value is  $\sim 0.3$. Moreover, for extremely low-metallicity winds, \citet{kudritzki2002} showed that $\delta$ might reach much higher values than $0.3$.

  The main characteristic of the $\delta$-slow solution is its lower terminal velocities than the fast solution. This $\delta$-slow solution might explain the observed terminal velocities from the winds of A-type supergiants  \citep{cure2011} and B-hypergiants \citep{venero2024}.

  From an observational point of view, the latest generation of large telescopes has opened a wide range of possibilities for studying massive stars. Continuous superior instrumentation development and improving the quality of the spectroscopic data are essential to understand the nature of these objects better and their chemical and dynamic evolution. Moreover, supergiant stars present themselves as very promising distance indicators through the application of the wind-momentum luminosity relationship \citep[WLR,][]{kudritzki1995} and the flux-weighted gravity-luminosity relationship \citep[FGLR,][]{kudritzki2003}. 
  Hence, to fully understand the nature of massive stars, it is necessary to perform accurate analyses of quantitative spectroscopy on large samples of these objects to obtain their physical properties \citep{Simon-Diaz2020}. 
  
  The only way to derive information from massive stars and their winds is by decoding their emitted radiation. Quantitative spectroscopy can extract numerous properties of massive stars and their winds through detailed non-local thermodynamic equilibrium (NLTE) modelling of their atmospheres and winds, generating a synthetic stellar spectrum for comparison with observational line profiles. Therefore, it is essential to use a state-of-the-art NLTE code capable of predicting a synthetic spectrum for a star with specific stellar and wind parameters and inhomogeneities. Whilst current multi-object spectrographs are undoubtedly capable of producing extensive collections of spectra \citep[see, e.g.][]{evans2005,simon2011a}, accurately modelling their atmospheres and winds is an inherently complex task due to the presence of NLTE processes. This complexity demands the use of advanced stellar modelling codes such as {\sc Fastwind}  \citep{santolaya1997,puls2005,rivero2012}, {\sc Cmfgen}  \citep{hillier1987,hillier1998,hillier2001,hillier2012}, and {\sc PoWR} \citep{hamann1987,todt2015}.

Regarding analysis techniques, several alternatives have been proposed by different authors to minimise the subjective component present in the widely used 'by-eye' techniques by introducing objective, automatic, and fast methods. These methods include using genetic algorithms employed by various studies such as \citet{mokiem2005}, \citet{tramper2014}, \citet{hawcroft2021}, and \citet{brands2022}. Another approach is a grid-based method, used by \citet{lefever2010} and \citet{simon2011b}, or the principal components analysis (PCA) algorithm designed by \citet{urbaneja2008}.
Given the multi-dimensionality of the parameter space considered during the modelling process, the solution is not unique,\footnote{At least when clumping is included, and the spectral range is not wide enough.} and the existence of similarly acceptable solutions with different combinations of stellar parameters and/or abundances can be obtained. Indeed, correctly identifying the range of acceptable parameters and abundances is as essential as determining the best fitting \citep{Simon-Diaz2020}. 

In the pioneering work of \citet{ppk1986}, the authors fitted a set of hydrodynamical models with the law now known as the $\beta$-law, with three parameters: stellar radius, terminal velocity, and the $\beta$ parameter. Nowadays, the $\beta$ parameter is empirically used as input in these codes, with typical values of $\beta=0.7$ to $1.5$ (or even larger) found in O- and B-type stars \citep[][]{puls1996,kudritzki2000}. However, in the case of B- and A-type supergiant stars, the $\beta$ value shows a tendency towards higher values \citep[][among others]{lefever2007,markova2008,haucke2018,rivet2020,almeida2022}, and in some cases, values as high as 3 to 4 \citep{stahl1991,crowther2006}. These higher values, however, lack a physical and hydrodynamic justification, and their use is justified posteriorly by the quality of the line profile fits achieved  \citep{kudritzki2000}.

In this work, we present the following elements:
  \begin{itemize}
        \item[i)] A new grid called the grId of Stellar atmOSphere and hydrodynamiC modELs for massivE Stars (ISOSCELES). The novelty of this new grid is that it uses the hydrodynamic solutions instead of the $\beta$ approximation for the velocity field. The hydrodynamic calculations were obtained by the stationary code {\sc Hydwind} \citep{cure2004} within the framework of the m-CAK theory. We used these hydrodynamical solutions as input in the {\sc Fastwind} code. In addition, these hydrodynamic solutions consider both the fast and $\delta$-slow solutions.
        
        \item[ii)]A new automatic grid-based procedure was implemented over ISOSCELES, specifically designed to study OBA-type stars in the optical and infrared range. With this tool, we expect to derive stellar ($T_{\rm{eff}}$, $\log \mathrm{g}$), wind ($v_{\infty}$, $\dot{M}$), and line-force ($\alpha$, $k$, $\delta$) parameters. 
  \end{itemize}
  
  Our work is organised as follows: Section \ref{gridm} describes our grid of models (hydrodynamic and stellar atmosphere), and Section \ref{methodology} describes the methods for the spectral line fitting. The ongoing results are shown in Section \ref{results}, and, finally, the discussion and conclusions of our work are given in Section \ref{disconc}.


\section{ISOSCELES grid}

\label{gridm}
ISOSCELES is the first grid involving the m-CAK hydrodynamics (instead of the generally used $\beta$-law\footnote{PoWR can use a double $\beta$-law, and WM-Basic uses an m-CAK fast solution.}) and NLTE radiative transport to generate synthetic line profiles for massive stars. ISOSCELES has been set up to cover almost the complete parameter space of OBA-type stars with solar metallicity, neglecting the effects of the stellar rotation.\footnote{To compare an observed line profile with synthetic ones, these should be broadened (convolved) with the respective stellar rotational and macroturbulence velocities.} 

The winds of massive stars are often modelled using a clumping factor, a parameter that accounts for small-scale density inhomogeneities (or `clumps') within the wind, rather than assuming a smooth, homogeneous outflow. In the current implementation of \textsc{ISOSCELES}, we adopted a clumping factor of unity (i.e. unclumped winds) across the entire grid. This choice allowed for a significant reduction in computational complexity when constructing millions of models; however, it also introduces notable limitations in the determination of wind parameters. In particular, density-squared diagnostics such as H$\alpha$ may lead to overestimated mass-loss rates, since a clumped wind with the same emission profile would require a lower average mass-loss rate. Moreover, the exclusion of clumping neglects porosity and optical depth effects that can affect both optical and UV line profiles. As a result, the current grid may not fully capture the impact of wind inhomogeneities in stars with stronger outflows. Future versions of \textsc{ISOSCELES} will incorporate a parameterised treatment of optically thin clumping to better constrain wind parameters and reduce degeneracies.

To produce this grid of synthetic line profiles with the code \textsc{Fastwind} (version 10.1.7), we first computed a grid of hydrodynamic wind solutions with the stationary code \textsc{Hydwind} for both m-CAK spherical solutions (fast and $\delta$-slow). \textsc{Fastwind} modified the lower part of the velocity profile from \textsc{Hydwind}, replacing the corresponding quasi-hydrostatic stratification to account for a better description of the velocity gradient (instead of using the Sobolev approximation).
The surface gravities range from $\log \mathrm{g}\,$=\,4.3 down to approximately 90\% of the Eddington limit, with increments of 0.15 dex. We considered 58 effective temperature values, from $9\,000$ K to $45\,000$ K, using steps of 500 K below $30\,000$ K and steps of 1\,000 K above it. 
Note that roughly below $13\,000$ K, the density structure predicted by \textsc{Fastwind} becomes inaccurate because of missing \ion{Fe}{II} lines in the optical, leading to underestimated radiative acceleration and consequently underestimated $\log \mathrm{g}$ values. However, we have implemented ISOSCELES from $9\,000$ K to avoid observed stars too close to the grid's borders. 
For calculating the non-LTE stellar atmosphere models, we used micro-turbulent 
velocities of 8 km s$^{-1}$ ($T_{\rm{eff}} < 15\,000$ K), 10 km s$^{-1}$ ($15\,000\,\rm{K} \leq T_{\rm{eff}} < 20\,000$\,K), and 15 km s$^{-1}$ ($T_{\rm{eff}} \geq 20\,000$ K). For all synthetic line profiles (as calculated in the final formal integral of the {\sc{Fastwind}} code), micro-turbulent velocities of 1, 5, 10, 15, 20, and 25 km s$^{-1}$ were employed. 

\begin{figure}[h!]
        \includegraphics[width=\columnwidth]{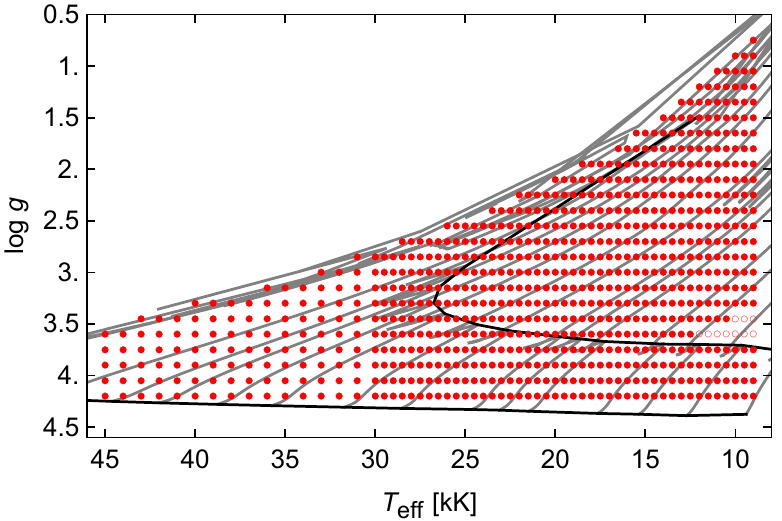}
        \caption{Location of ISOSCELES models in $T_{\rm{eff}}$--$\log\,\mathrm{g}$ plane (red dots). Grey solid lines represent the evolutionary tracks from $2 M_{\sun}$ to $60 M_{\sun}$ without rotation \citep{ekstrom2012}. Solid black lines correspond to both zero-age main sequence (ZAMS) and terminal-age main sequence (TAMS). Red (empty) circles indicate non-converged models in ISOSCELES.         
    \label{fullgrid}}
\end{figure}

In the current version of the grid, silicon (Si) is the only chemical element whose abundance is varied. We adopted five values of Si abundance, $\log \epsilon_\mathrm{Si} = 7.21, 7.36, 7.51$ (solar), 7.66, and 7.81, based on the solar reference from \citet{Asplund2009}. The choice to vary only Si at this stage is motivated by its strong diagnostic lines in the optical and near-infrared for B-type stars, such as \ion{Si}{iii} 4552\,\AA\ and \ion{Si}{iv} 4212\,\AA, which are key for determining $T_\mathrm{eff}$ in supergiants and early main-sequence stars. The abundances of all other species are set to the solar values, and the He abundance is fixed at He/H = 0.10. Future updates of ISOSCELES will be made available to us, including variations in additional elements (e.g. C, N, O) to enhance its diagnostic power and applicability to different metallicities and evolutionary phases.
Thus, the grid points were selected to cover the region of the $T_{\rm{eff}}$--$\log\,\mathrm{g}$ plane as shown in Fig. \ref{fullgrid}, where massive stars are located from the main sequence up to the blue supergiant phase. In this figure, each dot is further described by the other five parameters (see details in section \ref{descriptionISOSCELES}).

\cite{venero2024} showed that the winds of some B supergiants might be described by the $\delta$-slow solution. Therefore, we increased the number of models below $30\,000$ K to allow a more detailed analysis of the winds of BA supergiants.

\subsection{Grid of hydrodynamic models}
We computed a grid of hydrodynamic wind solutions with the code \textsc{Hydwind}. These wind solutions are then input into the \textsc{Fastwind} code. Each \textsc{Hydwind} model is described by the stellar parameters ($T_{\rm{eff}}$, $\log\,\mathrm{g}$, $R_{*}$) and the line force parameters from the m-CAK theory ($\alpha$, $k$, and $\delta$). All these spherically symmetric models are calculated without rotation and satisfy the boundary condition $\tau=2/3$ at the stellar surface for the optical depth $\tau$. For each given pair ($T_{\rm{eff}}$, $\log\,\mathrm{g}$), the radius was calculated using the FGLR \citep{kudritzki2003,kudritzki2008}, namely:

\begin{equation}
        M^{\rm{FGLR}}_{\rm{bol}} = 3.41 \times (\log \mathrm{g}_{\rm F} -1.5)-8.02\, ,
\end{equation} 

\noindent where $M^{\rm{FGLR}}_{\rm{bol}}$ is the bolometric magnitude, and $\log \mathrm{g}_{\rm F}=\log \mathrm{g} - 4 \log(T_{\rm eff}/10^4)$. 

While FGLR was originally calibrated for supergiant stars, we applied this relationship more broadly to estimate stellar radii for all stars in the grid, including those in lower luminosity classes. This approximation is adopted to ensure a uniform and computationally consistent method across the entire grid. However, we acknowledge that the FGLR may not be strictly valid outside the supergiant regime, and the derived radii for dwarfs should be interpreted as internal scaling parameters within the grid rather than precise physical values. For quantitative spectroscopic analyses of individual stars, these radii can be later rescaled using observed magnitudes, distances, and extinction, as discussed in Section \ref{descriptionISOSCELES} (see below) and in \citet{lefever2010}. In future updates, we plan to incorporate evolutionary models with appropriate luminosity-dependent radius estimates for each class.

The range of the line-force parameters is shown in Table~\ref{tab-hyd}, where the values of $\delta$ determine fast ($\delta \lesssim 0.24$) or $\delta$-slow solutions ($\delta \gtrsim 0.28$). Due to numerical problems, not all combinations of these parameters converge to a physical stationary hydrodynamic solution \cite[see, e.g.][]{venero2016}. As $\delta$-slow solutions converge less frequently than fast solutions, we used a denser grid of $\delta$ values from $\delta \gtrsim 0.26$.\\

\begin{figure}[h!]
        \includegraphics[width=\columnwidth]{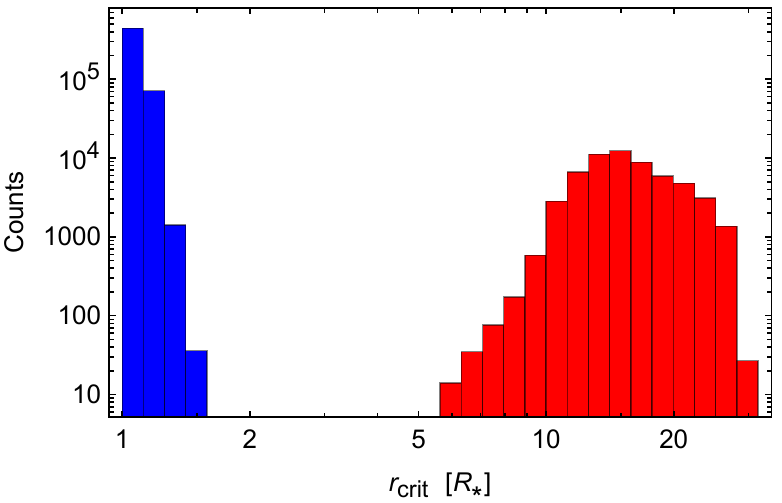}
        \caption{Histogram showing $r_{\rm{crit}}$ values from converged solutions. Blue bars represent fast solutions, and red bars correspond to $\delta$-slow solutions. Both axes are on a logarithmic scale. 
                \label{rcrit_plot}}
\end{figure}

Topologically, fast and $\delta$-slow solutions differ significantly in the singular (or critical) point location \citep{cure2023}. The fast solution has this point near the stellar surface, while the $\delta$-slow solution has a singular point very far from this surface. This property is illustrated in Fig. \ref{rcrit_plot}, which presents a histogram of the locations of the singular points ($r_{\rm{crit}}$) derived from the m-CAK equation of motion for the converged (fast and $\delta$-slow) solutions.  

\begin{table}
\caption{Combinations of line-force parameters for the grid of hydrodynamic models.}    
\label{tab-hyd}     
\centering                          
\begin{tabular}{c l}        
\hline\hline                
Parameter & Values\\    
\hline                        
                $\alpha$ & 0.45, 0.47, 0.51, 0.53, 0.55, 0.57, 0.61, 0.65 \\
                $k$    &  0.05 to 0.60 (step of 0.05)\\ 
                $\delta$    &  0.00,  0.04,  0.10,  0.14,  0.2, 0.24, \\
  & and 0.26 to 0.35 (step of 0.01) \\
\hline                                  
\end{tabular}
\end{table}

\subsection{ISOSCELES: Grid of stellar atmosphere models}
\label{descriptionISOSCELES}
A crucial aspect of analysing a massive star's spectra involves creating a grid of synthetic spectral lines and comparing it to the observed spectra by reproducing their different features. Because of its high computational efficiency, we used the stellar atmosphere code {\sc Fastwind}. This code considers non-LTE effects in spherical symmetry, with an explicit treatment of the stellar wind effects such as line blanketing/blocking and clumping. We used a version that enables the hydrodynamic wind solution from {\sc Hydwind}, ensuring a smooth transition between the pseudo-static photosphere and the lower wind layers. The transition point is located at $v = 0.1 v_{\rm{sound}}$, where $v_{\rm{sound}}$ is the sound speed. \\

The main advantage of {\sc Fastwind} concerning other similar codes is the possibility of generating realistic models in a short time interval, a crucial point for building a large dataset of synthetic line profiles. {\sc Fastwind} provides, in the optical and infrared wavelength ranges, H and He lines by default and lines from other elements, in dependence on purpose and available atomic data, in our case from Si. Using some of these lines, it is possible to determine the star's photospheric properties and the characteristics of its wind. In the ISOSCELES grid, each {\sc Fastwind} model is described by the following eight main parameters, $T_{\rm{eff}}$, $\log\,\mathrm{g}$, $R_{*}$, $\alpha$, $k$, $\delta$, $\log \epsilon_{\rm{Si}}$, and $v_{\rm{mic}}$, corresponding to effective temperature, the logarithm of the surface gravity, stellar radius, line-force parameters ($\alpha$, k and $\delta$), the logarithm of the Si abundance, and the micro-turbulence velocity, respectively.

Regarding the radius, we must remember that a rescaling to the real value is required once concrete objects are analysed. The actual radius should be determined from the visual magnitude, the star's distance, and reddening \citep{lefever2010}. 

In the end, $3\,440\,598$ synthetic spectra models (considering the six micro-turbulent velocities) were obtained, each containing 56 line profiles; these lines are tabulated in Appendix \ref{appendixA}.
Figure \ref{fullgrid} shows models with solutions (red dots) in the ($T_{\rm{eff}}$, $\log\,\mathrm{g}$) space. The grid was transformed to a binary format (FITS files) to reduce size. Currently, it has a size of 289 GB for the entire grid.

It is important to note that our method is not fully self-consistent since the radiation-wind interaction goes both ways. In principle, the wind structure and the spectrum must be obtained self-consistently through an iterative process. In addition, the m-CAK theory is considered a good approximation in the supersonic parts of the wind, but a questionable assumption in the sonic and sub-sonic portions due to the usage of the Sobolev approximation (e.g. \citealt{vink2022} and references therein). For some self-consistent works avoiding this approximation, see, for example, \citet{sander2017}, \citet{krticka2017}, \citet{gormaz2019}, and \citet{bjorklund2021}. However, despite these drawbacks, the results obtained from the ISOSCELES grid should have a better justification than the currently and widely calculated models with a $\beta$-law in quantitative spectroscopic analyses. Also, the ISOSCELES grid is helping us to clarify if the $\delta$-slow solution is a good candidate to solve some discrepancies, for example between predicted and observed mass-loss rates or the high values of $\beta$ obtained for some velocity laws.

Another assumption considered in our models is a smooth wind, i.e., a wind without clumps (regions of overdensity). These regions could become optically thick (especially in strong resonance lines), leading to porosity effects that may affect the line profiles and the mass-loss rate measurements. Similarly, optically thin clumps affect the mass-loss rate measurement from lines depending on $\rho^2$. Then, wind diagnostic lines that ISOSCELES cannot reproduce could be potential targets of interest for detailed investigations considering clumpy winds.

While the use of \textsc{Hydwind} velocity fields instead of a prescribed $\beta$-law might mitigate certain degeneracies in stellar and wind parameter determination, the exclusion of clumping as a free parameter introduces others. In particular, line diagnostics sensitive to $\rho^2$ (e.g. H$\alpha$) can be reproduced with different combinations of mass-loss rate and clumping factor, making it difficult to uniquely constrain $\dot{M}$ in the absence of additional observables or assumptions.

\section{Quantitative analysis procedure}
\label{methodology}

To identify the optimal synthetic model for each star under analysis, we developed a semi-automatic code that analyses a set of different observational line profiles for hydrogen, helium, and silicon. Specific lines of these elements can be selected to extract key stellar parameters and information about the wind parameters.

Each selected line contributes differently to the determination of stellar and wind parameters. In general, the Balmer lines (H$\gamma$ and H$\delta$) are primarily used to constrain the surface gravity. The H$\alpha$ profile is highly sensitive to the mass-loss rate ($\dot{M}$) due to its strong dependence on the wind density.

Different helium ionisation states are good indicators to estimate the effective temperature values: \ion{He}{I} 4471\,\AA\ and 6678\,\AA\ increase in strength towards cooler temperatures, while \ion{He}{II} 4541\,\AA\ and 4686\,\AA\ are stronger in hotter stars and can also serve as wind indicators in the presence of emission components. Similarly, the degree of ionisation of the silicon lines such as \ion{Si}{II} 4130\,\AA,\, \ion{Si}{III} 4552\,\AA,\ and \ion{Si}{IV} 4212\,\AA\ helps constrain $T_{\rm eff}$. This analysis is sensitive to silicon abundance variations, accounted for in the ISOSCELES grid. The combined fitting of these lines allows for a robust determination of the stellar parameters across a wide range of spectral types. The code additionally requires the rotational parameter, $v \sin i$, and the macroturbulence velocity, $v_{\rm{macro,}}$ as inputs, which were previously obtained with the {\sc iacob-broad}\footnote{http://research.iac.es/proyecto/iacob/pages/en/useful-tools.php} tool \citep{simon-diaz2014}. 

Optionally, the user can provide estimated ranges for the effective temperature, surface gravity, and solution type (fast or $\delta$-slow) to optimise the search procedure. If these optional parameters are not provided, the code uses effective temperatures from the Simbad astronomical database\footnote{\url{https://simbad.cds.unistra.fr/simbad/}} to constrain the search process or uses the entire grid if there is no information. On the other hand, $\log g$ is exhaustively searched throughout the whole ISOSCELES grid. 

The code utilises multi-processing to enhance efficiency, leveraging all available processors to execute:
i) convolution routines on synthetic line profiles; ii) interpolate the convolved line profiles within a user-defined spectral window at the wavelength of the observed spectra; and iii) perform a $\chi^2$ test, namely
\begin{equation}
        \boldsymbol{\chi}^2 = \frac{1}{n_{\rm{lines}}} \sum^{n_{\rm{lines}}}_{j} \frac{1}{n_\nu}  \sum^{n_{\rm{\nu}}}_{i} \frac{(y_{\rm{i,j}}-y_{\mathrm{obs}})^2}{{\rm{\sigma}}^2_\mathrm{L}}, 
\end{equation}
where $n_{\rm{lines}}$ represents the number of spectral lines to be fitted, $y_{\rm{obs}}$ is the normalised observed flux,  $y_{i,j}$ is the flux of the convolved synthetic model, $\rm{\sigma_\mathrm{L}}$ is the inverse signal-to-noise ratio (S/N) of the observed spectrum, and $n_{\rm{\nu}}$ corresponds to the number of wavelengths from the observed spectral line. This $\chi^2$ test was adapted from \citet{holgado2018}.The code compares the observed spectral lines against ISOSCELES's convolved and interpolated synthetic line profiles.

Finally, the code compiles the $\chi^2$ results, sorts them in ascending order, and identifies the best-fitting model, defined by the lowest $\chi^2$ value, averaged with equal weighting across all the used line profiles.\footnote{In a future version of this code, the user will define the weights of each line.} The best model is then exported to a PDF file that includes a graphical comparison of the synthetic spectral lines with the observational ones. 

The uncertainties in the derived stellar and wind parameters are primarily determined by the resolution of the ISOSCELES, which is defined by the step size of each parameter within the grid. Here, we did not account for the implementation of determined systematic errors of \citet{holgado2018}; this study will be the scope of future work.
The database is constructed using discrete intervals for key parameters such as effective temperature, surface gravity, and the line-force parameters ($\alpha$, $k$, and $\delta$). As a result, any derived parameter's uncertainty corresponds to the interval's size between adjacent grid points.\footnote{This is true if no noise and/or systematic errors are considered.} For example, the step size for effective temperature ($T_{\rm eff}$) and surface gravity (log $g$) directly defines the range within which the best-fit values may vary. Although this approach provides a systematic and well-defined estimate of uncertainties, future improvements could involve interpolating between grid points or employing Bayesian methods to further refine the estimates and reduce parameter uncertainties.

To move beyond assigning uncertainties based solely on the grid resolution, we conducted an additional analysis, presented in Appendix~\ref{uncertainties}, to evaluate the sensitivity of the error estimates to the number of top-ranked models considered. Following a quantile-based method similar to that of \citet{Turis2025}, we tested different values of $N$ and found that the resulting uncertainties, particularly for wind parameters, depend strongly on this choice. This highlights the limitations of arbitrary model selection thresholds and motivates the future implementation of a more rigorous statistical framework.

As an additional robustness check, we also performed a separate temperature fitting using only helium and, alternatively, only silicon lines, while keeping the wind parameters fixed. Both diagnostics returned effective temperature values consistent within the estimated uncertainties. The details of this test are described in Appendix~\ref{uncertainties}.

\section{Quantitative spectroscopy}
\label{results}
To test the capabilities of the ISOSCELES database in accurately deriving stellar and wind parameters across a range of physical conditions, we selected six stars, a subset of three dwarf stars and another with three supergiant stars; each subset has a star that lies on the hot side of the bi-stability jump \citep[see, e.g.][hereafter BS-jump]{vink2022}. The selection of stars on both sides of the BS-jump allows us to test whether the ISOSCELES framework can account for the different wind behaviours associated with this transition. Information about the used observations (dates, SNR, and instruments) is summarised in Appendix \ref{obs}.

Considering that the mass-loss rate, $\dot{M}$, strongly determines the shape of the H$\alpha$ line profile, the selected stars show different shapes of this line, allowing us to test the performance of our ISOSCELES procedure for a wide range of $\dot{M}$.
We analysed a set of different spectral lines, including H$\alpha$, and contrasted the results with previous studies that use {\sc Cmfgen/Fastwind} codes with the $\beta$-law velocity profiles. 

\subsection{Analysis of dwarf stars}
\label{section-dwarf}
In this section, we test the ability of the ISOSCELES grid to model the spectra of dwarf stars. Since the wind of these stars is too weak, we only focus on the stellar parameters, $T_{\rm{eff}}$ and $\log \mathrm{g}$, testing the photospheric structure.
        
As the analysis of line profiles is highly sensitive to the region where the transition between the quasi-hydrostatic layers and the stellar wind occurs, which is very close to the stellar surface, it is crucial to assess whether the inclusion of m-CAK wind solutions affects the outer layers of the photosphere.
The selected dwarf stars representing different stellar types are HD 14633, HD 35299, and HD 35912.

\subsubsection{HD 14633} %
This is an ON8.5 V star, and the spectrum is from the IACOB\footnote{https://research.iac.es/proyecto/iacob/iacobcat/} database \citep{simon2011a}. Figure~\ref{HD14633} (see Appendix \ref{dwarf-fig}) shows nine spectral lines
and their best fast and $\delta$-slow fits, where the fast solution is the better one. Some fitted line profiles are commonly used as a criterion to determine stellar parameters, summarised in Table \ref{tab:HD14633}, where we compared our results with the ones from \citet{Aschenbrenner2023}.

The main differences between the observation and the synthetic line profiles are found in the core of the lines formed in the photosphere and the wind. The fast wind solution fits most of the line profiles better. Note that while \ion{He}{II} 4686\,\AA\ is well described, the core of H$\alpha$ is too shallow for the fast solution.

\begin{table}[h!]
\caption{Comparison of stellar parameters from fast and $\delta$-slow best-fit solutions for HD 14633, along with the analysis performed by \citet{Aschenbrenner2023}.
}
\label{tab:HD14633}
\centering
\begin{tabular}{lccc}
\hline\hline
HD 14633 & Fast & $\delta$-slow & \citeauthor{Aschenbrenner2023} \\
\hline
$T_{\mathrm{eff}}$ [K] & 36\,000 & 35\,000 & 34\,000 \\
$\log \mathrm{g}$ & 3.9 & 3.9 & 3.9 \\
$\log \epsilon_{\rm{Si}}$ & 7.51 & 7.36 & 7.41 \\
$v\sin i$ [km/s] & 121 & 121 & 125 \\
$v_{\rm{mic}}$ [km/s] & 20 & 20 & 6 \\
$v_{\rm{mac}}$ [km/s] & 69 & 69 & 74 \\
\hline
\end{tabular}
\end{table}

\subsubsection{HD 35299}
This second tested dwarf star is a B1.5 V from the IACOB database. Figure~\ref{HD35299} shows a set of nine spectral lines that are best suited for this spectral type \citep[see, e.g.][]{evans2015}. We compare our results (see Table \ref{tab:HD35299}) with the ones from \citet{nieva2012}.
Both wind regimes predict almost similar line profiles, implying that the mass-loss rate is very low.

\begin{table}[h!]
\caption{Same as Table~\ref{tab:HD14633}, but for HD 35299, compared with the results from \citet{nieva2012}.
}
\label{tab:HD35299}
\centering
\begin{tabular}{lccc}
\hline\hline
HD 35299 & Fast & $\delta$-slow & \citeauthor{nieva2012} \\
\hline
$T_{\mathrm{eff}}$ [K] & 23\,500 & 23\,500 & 23\,500 \\
$\log \mathrm{g}$ & 4.2 & 4.2 & 4.2 \\
$\log \epsilon_{\rm{Si}}$ & 7.51 & 7.51 & 7.56 \\
$v\sin i$ [km/s] & 4 & 4 & 8 \\
$v_{\rm{mic}}$ [km/s] & 1 & 1 & 0 \\
$v_{\rm{mac}}$ [km/s] & 15 & 15 & - \\
\hline
\end{tabular}
\end{table}

\subsubsection{HD 35912}
Our last dwarf star has a spectral type B2V; we also obtained an observed spectrum from the IACOB database. Figure~\ref{HD35912} and Table~\ref{tab:HD35912} summarise the results of the fitting procedure and a comparison with the work from \citet{nieva2011}. The predicted line profiles from both wind regimes exhibit a high degree of concordance.

\begin{table}[h!]
\caption{Same as Table~\ref{tab:HD14633}, but for HD 35912, compared with the results from \citet{nieva2011}.
}
\label{tab:HD35912}
\centering
\begin{tabular}{lccc}
\hline\hline
HD 35912 & Fast & $\delta$-slow & \citeauthor{nieva2011} \\
\hline
$T_{\mathrm{eff}}$ [K] & 19\,500 & 19\,500 & 19\,000 \\
$\log \mathrm{g}$ & 4.05 & 4.05 & 4.00 \\
$\log \epsilon_{\rm{Si}}$ & 7.51 & 7.36 & 7.50 \\
$v\sin i$ [km/s] & 11 & 11 & 15 \\
$v_{\rm{mic}}$ [km/s] & 1 & 1 & 2 \\
$v_{\rm{mac}}$ [km/s] & 20 & 20 & 8 \\
\hline
\end{tabular}
\end{table}

As expected, the solutions for absorption lines are very similar for both $\delta$-slow and fast regimes, indicating that the wind has minimal impact on these lines. Exceptions to this include H$\alpha$ and \ion{He}{II} 4686 \AA, where more pronounced differences are observed in the cores due to the `fill-in' effect caused by the tenuous wind.

As a final remark of this section, the photospheric structure of the ISOSCELES models computed using {\sc Hydwind} and {\sc Fastwind} delivers values of the stellar parameters close to published ones. This gives us confidence in fitting more complex atmospheres that include strong winds, as in the case of supergiant stars.

The successful fits obtained for the three dwarf stars are largely expected, given that their winds are extremely weak and have negligible influence on the line formation in the optical regime. Consequently, the hydrodynamic input from \textsc{Hydwind} plays a minor role, and the synthetic spectra are dominated by the NLTE atmospheric structure and radiative transfer handled by \textsc{Fastwind}. In this sense, our comparison with previous analyses based on {\sc DETAIL/SURFACE} or {\sc TLUSTY} models \citep[e.g.][]{nieva2011,nieva2012} serves more as a consistency check between independent codes rather than a full test of the hydrodynamic component of \textsc{ISOSCELES}. Nonetheless, this validation step is important to ensure that the grid can reliably recover stellar parameters for stars with negligible winds before moving on to more complex cases, such as supergiants, where the wind solution significantly affects the atmospheric structure and emergent spectrum.

\subsection{Analysis of supergiant stars}
In this section, we analyse three early-type supergiant stars, all with P-Cygni or emission H$\alpha$ line profiles, which are very sensitive to the mass-loss rate. Due to this feature, we added the wind parameters in addition to the stellar ones.
\subsubsection{HD 14947}
\begin{figure}[h!]
        \includegraphics[width=\columnwidth]{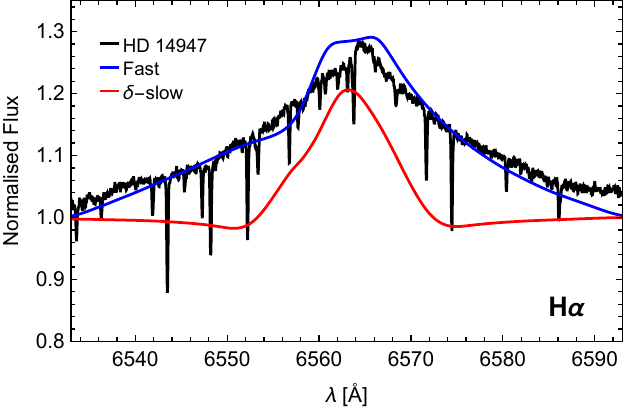}
        \caption{H$\alpha$ spectral-line fits for O-type star HD 14947. The black line represents the observed spectrum, while the best fit from the ISOSCELES database with a fast solution and $\delta$-slow solution are shown in blue and red, respectively. The fast regime provides a better fit to this line profile.
 \label{HD14947}}
\end{figure}

\begin{figure*}[h!]
\centering
   \includegraphics[width=17cm]{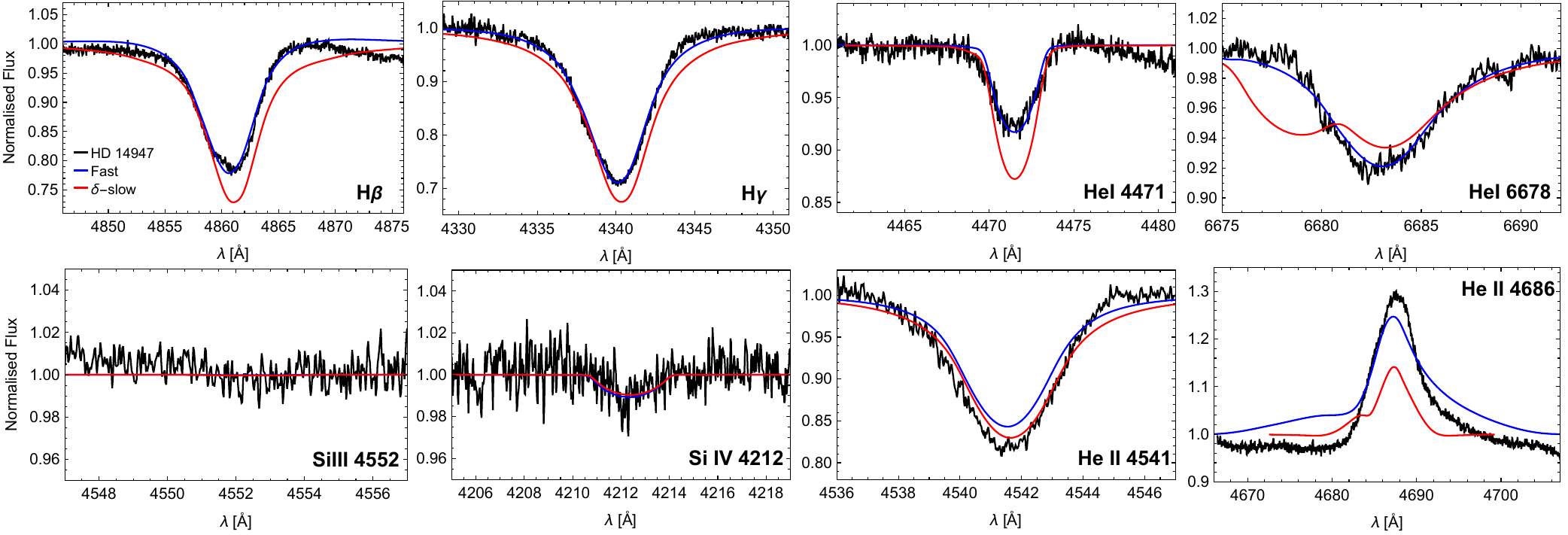}
        \caption{Best spectral-line fits for HD 14947. The black line depicts the observed spectrum. The best fit with a fast solution is shown in blue, while the best fit with a $\delta$-slow solution is in red. Clearly, the fast solution better describes the wind of this early-type star.
\label{HD14947-all}}
\end{figure*}

\begin{figure}[h!]          
\includegraphics[width=\columnwidth]{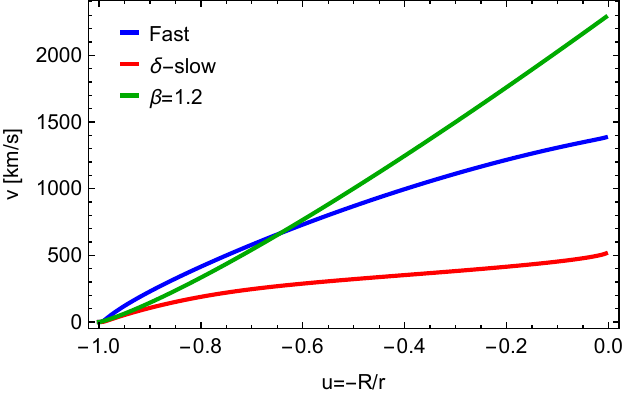}
        \caption{Velocity profiles as function of inverse radial coordinate $u$ for the best-fit solutions of HD 14947. The velocity profile for the fast solution is shown in blue, while the $\delta$-slow solution is depicted in red. Additionally, the green line represents the $\beta$-law velocity field, based on the parameters from \cite{Bouret2012}.
}
     \label{HD14947-vel}
\end{figure}

The star HD 14947, classified as spectral type O4.5 If \citep{sota2011}, was studied using spectroscopic data from the IACOB database. Figure~\ref{HD14947} presents the H$\alpha$ line profile,  while Fig.~\ref {HD14947-all} displays eight additional spectral lines. The results show that the fast solution yields a better overall fit to the observed profiles than the $\delta$-slow solution. However, it is important to note that the poor fit in the wings of the \ion{He}{II} 4686 \AA\ line suggests that the outer velocity field is not accurately reproduced. This discrepancy highlights the need for further refinement in modelling the wind structure using UV lines. The velocity profiles used as input for generating the synthetic spectra are displayed in Fig.~\ref{HD14947-vel}.

Table~\ref{tab:HD14947} shows stellar and wind parameters for both hydrodynamical solutions. For comparison purposes, in the last column, we add the results from \cite{Bouret2012}, which used a $\beta$-law description for the velocity field. It is worth noting that ISOSCELES is a grid of models, all with a clumping factor of one ($f_{cl}=1$), while the results from \cite{Bouret2012} consider an optically thin clumping factor with a value of $f_{cl}=33$ ($f_{\infty}=0.03$).

\begin{table}[h!]
\caption{Same as Table~\ref{tab:HD14633}, but for HD 14947, compared with the results from \citet{Bouret2012}, who adopted a $\beta$-law velocity profile.
}
\label{tab:HD14947}
\centering
\begin{tabular}{lccc}
\hline\hline
HD 14947 & Fast & $\delta$-slow & \citeauthor{Bouret2012} \\
\hline
$T_{\mathrm{eff}}$ [K] & 39\,000 & 39\,000 & 37\,000 \\
$\log \mathrm{g}$ & 3.75 & 3.9 & 3.52 \\
$\log \epsilon_{\rm{Si}}$ & 7.51 & 7.51 & 7.51 \\
$\alpha$ & 0.55 & 0.45 & - \\
$\kappa$ & 0.3 & 0.25 & - \\
$\delta$ & 0.1 & 0.3 & - \\
$\beta$ & - & - & 1.2 \\
$\dot{M}$ [$10^{-6} M_\odot$ / yr] & $2.4 $ & $0.42 $ & $1.41 $ \\
$v_\infty$ [km/s] & 1384.44 & 514.31 & 2300.0  \\
$v\sin i$ [km/s] & 114 & 114 & 130 \\
$v_{\rm{mic}}$ [km/s] & 5 & 10 & 15 \\
$v_{\rm{mac}}$ [km/s] & 22 & 22 & 36 \\
$f_{cl}$  & 1 & 1 & 33 \\
\hline
\end{tabular}
\tablefoot{Direct measurement from UV lines by \citet{prinja1990} determined a terminal velocity of 1885 [km/s].}
\end{table}

\subsubsection{HD 206165} 
\label{B2Ib-star}

\begin{figure}[h!]
        \includegraphics[width=\columnwidth]{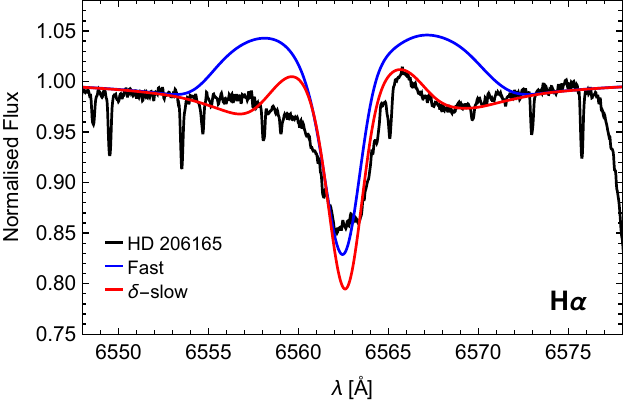}
        \caption{
    Same as Fig. \ref{HD14947}, but for HD 206165. Here, the $\delta$-slow solution is the one that best fits the observed line profile.
    \label{HD206165}}
\end{figure}

\begin{figure*}[h!]
\centering
        \includegraphics[width=17cm]{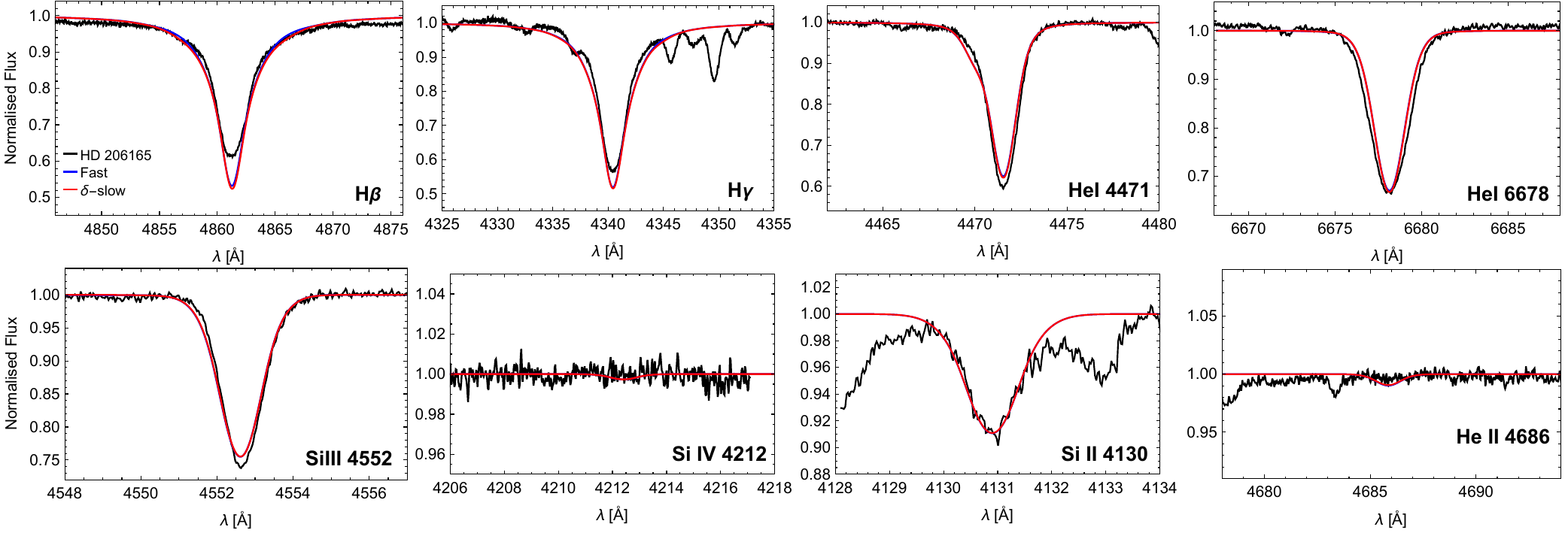}
        \caption{
    Same as Fig. \ref{HD14947-all}, but for HD 206165. Both solutions show the same behaviour, demonstrating that these lines are not as sensitive to the mass-loss rate and the velocity profile as H$\alpha$.
    \label{HD206165-all}}
\end{figure*}

\begin{figure}[h!]         
\includegraphics[width=\columnwidth]{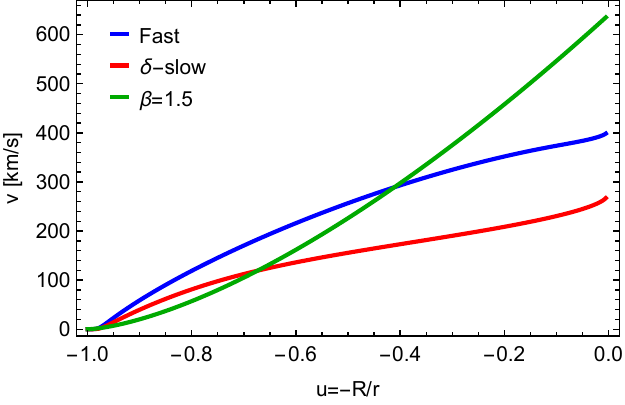}
        \caption{Same as Fig.~\ref{HD14947-vel}, but for HD 206165. The $\beta$ velocity profile is from the analysis of \cite{markova2008}.
    \label{HD206165-vel}}
\end{figure}

The star HD 206165, classified as spectral type B2 Ib \citep{negueruela2024}, was analysed using spectroscopic observations from the IACOB database. Figure \ref{HD206165} shows the H$\alpha$ line profile, where the $\delta$-slow solution fits this line profile much better than the fast one. Eight additional lines are shown in Fig. \ref{HD206165-all}. The synthetic lines from both wind solutions are indistinguishable, suggesting that H$\alpha$ is the only sensitive line in our set of optical line profiles to distinguish the different wind regimes. Fig.~\ref{HD206165-vel} shows the corresponding velocity profiles. Table~\ref{tab:HD206165b} summarises the stellar and wind parameters from our fits and the results from the work of \citet{markova2008}, which used a $\beta$-law to describe the velocity field.

\begin{table}[h!]
\caption{Same as Table~\ref{tab:HD14633}, but for HD 206165, compared with the results from \cite{markova2008}, who adopted a $\beta$-law velocity profile.
}
\label{tab:HD206165b}
\centering
\begin{tabular}{lccc}
\hline\hline
HD 206165 & Fast & $\delta$-slow & \citeauthor{markova2008} \\
\hline
$T_{\mathrm{eff}}$ [K]   & 20\,000     & 20\,000   &  19\,300   \\
$\log \mathrm{g}$       & 2.7         & 2.7   &    2.51     \\
$\log \epsilon_{\rm{Si}}$ & 7.51         &  7.51  &     7.58    \\
$\alpha$       & 0.45        & 0.45  &     -      \\
$k$            & 0.3         & 0.3  &     -     \\
$\delta$       & 0.2        & 0.3  &     -      \\
$\beta$         & -         & -        & 1.5  \\
$\dot{M}$ [$10^{-6} M_\odot$/yr] & $0.175$ & $0.871$ & $0.269$ \\
$v_\infty$ [km/s] & 398.55   & 267.13    &  640  \\ 
$v_{\rm{mic}}$ [km/s] & 10 & 10 & 8 \\
$v \sin i$ [km/s] & 42 & 42 & 45 \\
$v_{\rm{mac}}$ [km/s] & 63 & 63 & 57 \\
$f_{cl}$  & 1 & 1 & 1 \\
\hline
\end{tabular}
\tablefoot{\citet{Bernini-Peron2023} reported a $v_\infty=900$ [km/s] from UV lines.}
\end{table}

\subsubsection{HD 53138} 
\label{B-star}
\begin{figure}[h!]
        \includegraphics[width=\columnwidth]{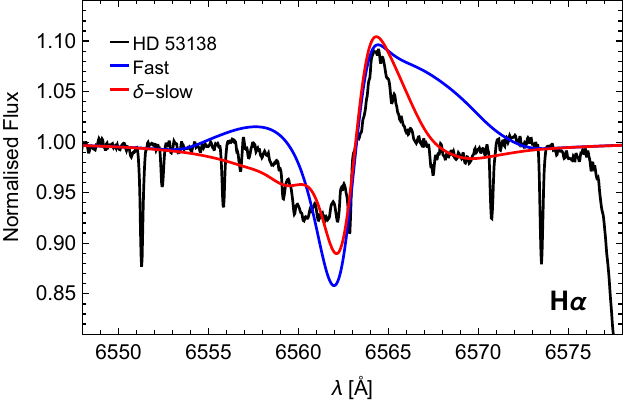}
        \caption{ Same as Fig. \ref{HD14947}, but for HD 53138. In this case, the $\delta$-slow solution is the one that best fits the observed line profile.
    \label{HD53138b}}
\end{figure}

\begin{figure*}[h!]
\centering
   \includegraphics[width=17cm]{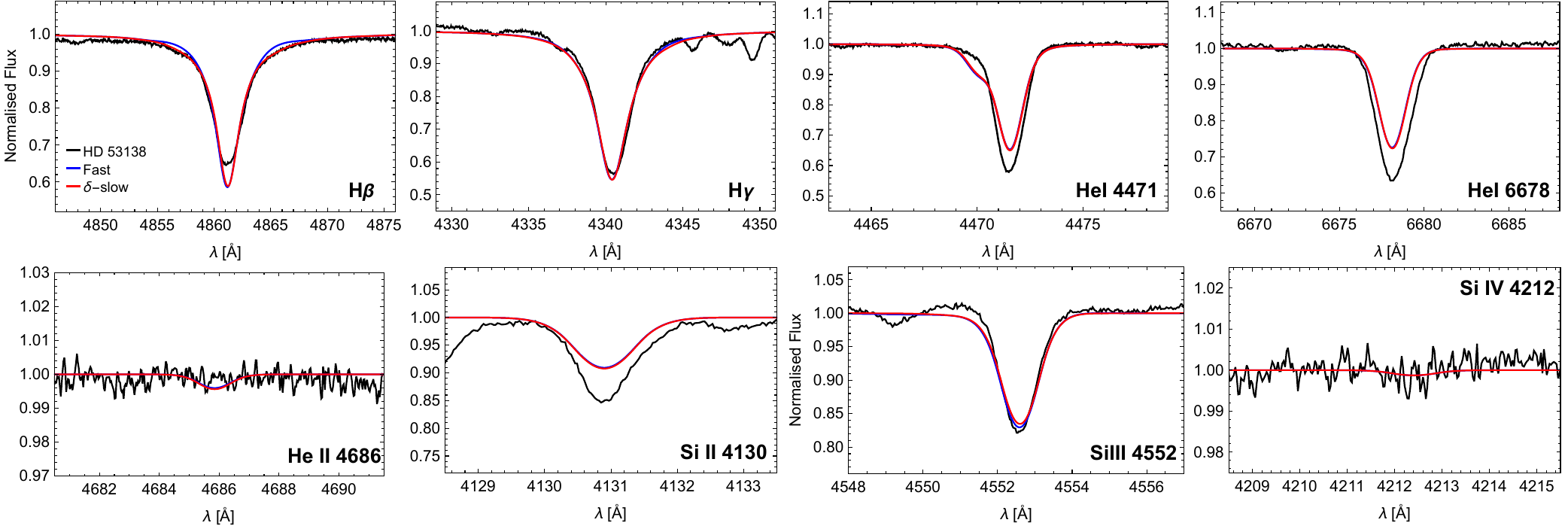}
        \caption{Best spectral-line fits for HD 53138. The black line depicts the observed spectrum. The best fit with a fast solution is shown in blue, while the best fit with a $\delta$-slow solution is in red.
    \label{HD53138-all}}
\end{figure*}

The star HD 53138 is a blue supergiant of spectral type B3 Ia \citep{negueruela2024}. Figure \ref{HD53138b} shows the H$\alpha$ line profile, and Figure \ref{HD53138-all} shows the other eight available lines from our spectrum used in the fitting procedure. The $\delta$-slow solution fits the observed profile better than the fast solution, mainly because the red wing of the H$\alpha$ line has a better agreement with the observation. 

In Table \ref{tab:HD53138b}, some stellar and wind parameters for both hydrodynamical solutions are tabulated. For comparison purposes, in the last column, we added the results from \cite{haucke2018}, which used a $\beta$-law description for the velocity field.

\begin{figure}[h!]         
\includegraphics[width=\columnwidth]{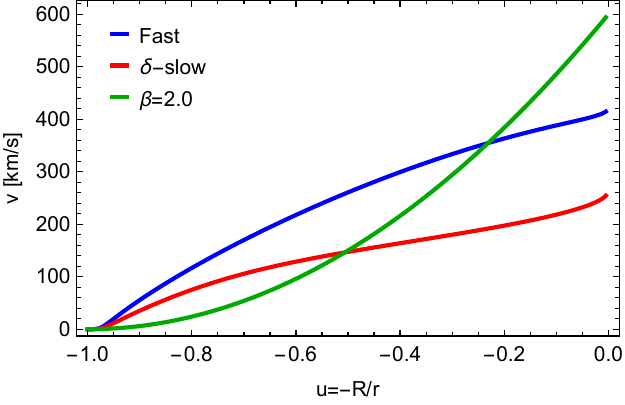}
        \caption{Same as Fig.~\ref{HD14947-vel}, but for HD 53138. The $\beta$ velocity profile is from the analysis of \cite{haucke2018}. 
    }
\end{figure}

\begin{table}[h!]
\caption{Same as Table~\ref{tab:HD14633}, but for HD 53138, compared with the results from \cite{haucke2018}, who adopted a $\beta$-law velocity profile.}
\label{tab:HD53138b}
\centering
\begin{tabular}{lccc}
\hline\hline
HD 53138 & Fast & $\delta$-slow & \citeauthor{haucke2018} \\
\hline
$T_{\mathrm{eff}}$ [K]   & 18\,500     & 18\,500   &  18\,000   \\
$\log \mathrm{g}$       & 2.4         & 2.4   &    2.25     \\
$\log \epsilon_{\rm{Si}}$ & 7.51         &  7.51  &   7.51      \\
$\alpha$       & 0.53        & 0.53  &     -      \\
$k$            & 0.15         & 0.15  &     -     \\
$\delta$       & 0.20        & 0.34  &     -      \\
$\beta$         & -         & -        & 2.0  \\
$\dot{M}$ [$10^{-6} M_\odot$/yr] & $0.449$ & $0.244$ & $ 0.24 $ \\
$v_\infty$ [km/s] & 414.66   & 254.31    &  600  \\ 
$v_{\rm{mic}}$ [km/s] & 5 & 5 & 10 \\
$v \sin i$ [km/s] & 41 & 41 & 40 \\
$v_{\rm{mac}}$ [km/s] & 55 & 55 & 50 \\
$f_{cl}$  & 1 & 1 & 1 \\
\hline
\end{tabular}
\tablefoot{Using only H$\alpha$ line profile fitting, a value for the terminal velocity of  212.2 [km/s] for the $\delta$-slow solution was estimated by \citet{venero2024}. \citet{Bernini-Peron2023} reported a $v_\infty=680$ [km/s] from UV lines.}
\end{table}

The standard fitting procedure for massive star winds typically relies on the empirical $\beta$-law for the velocity field and allows for different values of the clumping factor, introducing degeneracies between the derived mass-loss rate and terminal velocity. In this work, we adopted a different approach: we employed self-consistent hydrodynamic m-CAK solutions (both fast and $\delta$-slow) for the wind-velocity structure, but we fixed the clumping factor to $f_{\text{cl}} = 1$ (i.e. assuming unclumped winds). This strategy significantly reduces the degeneracy between $\dot{M}$ and $v_\infty$ by eliminating the additional free parameters associated with the $\beta$-law and variable clumping, allowing for a more physically motivated determination of wind parameters.
Nevertheless, it is important to note that using a single value of the clumping factor is a simplification; real stellar winds are expected to be inhomogeneous, and future works will need to explore a range of $f_{\text{cl}}$ values to fully capture the impact of clumping on wind diagnostics. While optical lines such as H$\alpha$ remain sensitive to both $\dot{M}$ and $v_\infty$, UV resonance lines provide crucial constraints on $v_\infty$ and the wind's ionisation structure, helping to break these degeneracies. The comparison with recent UV-based studies \citep[e.g.][for HD~206165 and HD~53138]{Bernini-Peron2023} highlights the added value of UV diagnostics, which often yield higher terminal velocities than those inferred from optical-only, $\delta$-slow fits.
This motivates our planned extension of \textsc{ISOSCELES} to include UV synthetic spectra and integrated optical+UV fitting routines. By combining hydrodynamic wind solutions with a more comprehensive treatment of clumping in future work, we aim to minimise degeneracies further and improve the reliability of derived stellar wind parameters.

As initially motivated, the selection of the six targets was designed to sample stars on both sides of the so-called bi-stability jump (BS-jump), a regime where significant changes in wind properties, particularly terminal velocity and mass-loss rate, are observed near $T_{\rm eff} \sim 21\,000$–$25\,000$\,K \citep[see, e.g.][]{vink2022}. Interestingly, our results show that the $\delta$-slow hydrodynamic solution systematically provides better fits to the cooler stars below the BS-jump, such as HD~206165 and HD~53138, while hotter stars such as HD~14947 are better described by the fast solution. This pattern suggests that ISOSCELES may offer a physically motivated mechanism, within the m-CAK theory, for interpreting the BS-jump as a bifurcation between distinct wind regimes. While this result alone does not constitute a predictive proof of the BS-jump, it supports the idea that the $\delta$-slow solution could provide a consistent hydrodynamic explanation for this observational phenomenon. Further investigation is warranted using a larger sample and, ideally, the inclusion of UV diagnostics to evaluate whether the $\delta$-slow/fast dichotomy aligns consistently with the BS-jump across the HR diagram.

\section{Discussion and conclusions\label{disconc}}

The ISOSCELES database, introduced in this study, represents an advancement in the quantitative spectroscopic analysis of massive stars. We propose that by incorporating hydrodynamic wind solutions derived from the m-CAK theory, instead of relying solely on the conventional $\beta$-law approximation, we provide a more physically robust framework for modelling stellar winds. Including fast and $\delta$-slow solutions within ISOSCELES potentially permits a more precise representation of wind dynamics. This is particularly relevant for addressing previously observed discrepancies, specifically concerning terminal velocities and mass-loss rates in B-supergiants, where the $\delta$-slow solutions might offer a better fit. However, multi-wavelength analyses and a larger sample of stars are still necessary to draw a definitive conclusion. We plan to address this in future work to maximise the potential of our approach.

In applying ISOSCELES to a limited, yet carefully chosen, sample of dwarf stars (HD 14633, HD 35299, and HD 35912) and supergiant stars (HD\,14947, HD\,206165, and HD\,53138), we have observed its ability to reproduce their optical spectral profiles with reasonable fidelity. For the O-type star HD\,14947, the fast solution delivered a superior fit to the H$\alpha$ line profile, aligning with expectations for stars possessing high terminal velocities. This is consistent with previous studies and provides a degree of validation for the fast solution within the ISOSCELES framework. Conversely, for the two B-type supergiant analysed, the $\delta$-slow solution demonstrated a better fit, particularly in the H$\alpha$ profile. This underscores the potential of the $\delta$-slow solution in modelling stars with lower terminal velocities on the order of a few hundred km/s and showcases the advantages of ISOSCELES. These findings, while promising, are preliminary, and we agree on the need for further refinement and expansion of the ISOSCELES database. 

The derived $\alpha$ values from our best-fit models fall within the expected theoretical range for line-force parameters. However, for B-supergiants, the retrieved values are slightly higher than those predicted by theoretical studies, such as those presented by \citet{puls2000}. This deviation may reflect the influence of additional physical processes not fully captured in current hydrodynamic models. This result underscores the importance of further investigation into the parametrisation of radiative acceleration. It suggests that improved empirical constraints alongside refined theoretical modelling will be crucial for advancing our understanding of radiation-driven winds in this regime.

Given the computational expense of incorporating clumping, the assumption of smooth, unclumped winds in our models is an (currently) inevitable/forced simplification. However, we acknowledge that clumping in actual stellar winds can influence the observed spectra and derived parameters, particularly affecting H$\alpha$ and UV lines, leading to overestimates of mass-loss rates \citep[see, e.g.][]{hawcroft2021}.  

Similarly, while computationally expedient, the dependence on the Sobolev approximation may not fully encapsulate the intricate physics of the sonic and subsonic wind regions. In these regions, the velocity field is either strongly curved or its gradient is slight, and the Sobolev approximation can break down. However, as previously noted, this limitation does not affect our analysis, as we employed an alternative approach for calculating the wind structure that does not depend on the Sobolev approximation in these regions.

Furthermore, although non-solar metallicity environments are crucial for understanding stellar evolution across different galaxies, they have not yet been explored in this study. Applying our database to other stellar populations, such as those in the Magellanic Clouds or other nearby galaxies, will require the development of additional grids with appropriately adjusted metallicities. We plan to develop a more comprehensive grid of models, including an optimised $\chi^2$ fitting procedure and, where justified, a finer sampling of $\delta$ values. This approach will enable a more robust and quantitative exploration of the parameter space than is possible through visual comparison alone. 
 
Incorporating grids for supergiants with varying metallicities is essential to fully leverage the X-Shooting ULLYSES project \citep{vink2023}. This will enhance our ability to constrain stellar and wind parameters with greater precision, ultimately improving our understanding of massive stars in diverse galactic environments.

It is worth emphasising that ISOSCELES is still a developing project with the potential to evolve in several important directions. One key avenue for improvement involves refining the physics incorporated into the grid generation. This includes addressing current limitations, such as the treatment of clumping and the reliance on the Sobolev approximation in the wind calculations, both of which can impact the accuracy of wind parameter estimates. Additionally, a more precise description of the velocity field could be achieved by incorporating a detailed treatment of line overlaps, as proposed by \citet{poniatowski2022}. Their iterative procedure self-consistently accounts for the complex interactions between radiation and matter, providing a more accurate representation of the wind dynamics and radiation field coupling. Implementing such enhancements would further improve the reliability of ISOSCELES in modelling massive star winds and interpreting observational data.

Secondly, we could explore alternative techniques for fitting observed spectra, such as employing machine learning algorithms \citep[see, e.g.][]{olivares2024} or Bayesian inference methods to enhance parameter estimation and provide a more robust quantification of uncertainties. Another promising approach is to deconvolve the observed line profile to that of a non-rotating star \citep[see, e.g.][]{escarate2023}. This would eliminate the need to convolve all synthetic profiles during the comparison process, significantly reducing computational time and resource requirements.

Crucially, we plan to extend the ISOSCELES analysis to include the UV spectral range, which is rich in diagnostic information about the winds of massive stars. The UV spectrum contains numerous resonance lines that are highly sensitive to the wind density and velocity structure. By simultaneously fitting spectral features in both the optical and UV ranges, we can achieve much tighter constraints on key wind parameters, including the terminal velocity, mass-loss rate, and the parameters governing the velocity law. This multi-wavelength approach is essential for breaking degeneracies that arise when relying solely on optical lines, leading to a more comprehensive and robust characterisation of the stellar wind. This enhancement will be integrated into the ISOSCELES database, improving the accuracy of derived physical and wind parameters.

One of our future objectives is to incorporate CNO elements into the ISOSCELES framework to further enhance its precision in parameter derivation. UV data are critical as they provide firm constraints on the properties of stellar winds. The UV range encompasses many important spectral lines, including resonance lines, intercombination lines, and subordinate lines. Resonance lines, in particular, are invaluable for constraining low mass-loss rates (of the order of $\sim 10^{-9} M_\odot/\text{yr}$), where the H$\alpha$ line becomes insensitive \citep{hilier2020}.

The synergy between ISOSCELES and UV spectra will offer a significant potential for advancing our understanding of mass loss in B supergiants. Another key advantage of UV lines, especially those involving resonance transitions and large optical depths, is their ability to provide precise measurements of terminal wind velocities. As a result, incorporating the UV regime will be critical for obtaining accurate and reliable information on terminal velocities and other wind properties, thereby enhancing the overall effectiveness of our spectroscopic models. However, to fully exploit the diagnostic power of the UV, it is essential to account for optically thick clumping.

In conclusion, we present the ISOSCELES database as an evolving and valuable resource designed to benefit the scientific community, especially with its planned extension to the UV spectral range. While we acknowledge its limitations, we are optimistic that ISOSCELES will contribute significantly to our understanding of massive stars and their role in galactic evolution with ongoing development. Its ability to handle a wide range of stellar parameters, along with its potential for expansion to different metallicities and wavelength ranges, positions it as a promising tool for analysing large datasets from current and future observational facilities, such as the ELT and JWST.

We believe ISOSCELES will be instrumental in interpreting the vast amounts of data expected from these next-generation instruments, and we aim to enhance its accuracy and efficiency over the coming years. To foster collaboration and support further advancements in the field, we welcome community engagement and are open to collaborations upon reasonable request. We also plan to make the ISOSCELES database available in the future.

\begin{acknowledgements}
The authors thank Joachim Puls for his valuable comments and insights on this manuscript.
We also thank the anonymous referee for the valuable comments and suggestions, which helped to improve the clarity and quality of this manuscript.
The authors are grateful for support from ANID / FONDO 2023 ALMA / 31230039.  MC, IA \& CA thank the support from ANID FONDECYT project 1230131. MC and CA acknowledge partial support from Centro de Astrofísica de Valparaíso. NM thanks the support from ANID BECAS / DOCTORADO NACIONAL 21221364. ROJV acknowledges financial support from CONICET (PIP 11220200101337CO) and the Universidad Nacional de La Plata (Programa de Incentivos 11/G192 and 11/G193). This work has been possible thanks to the use of AWS-U.Chile-NLHPC credits. Powered@NLHPC: This research was partially supported by the NLHPC's supercomputing infrastructure (ECM-02). This project has received funding from the European Union’s Framework Programme for Research and Innovation, Horizon 2020 (2014-2020), under the Marie Skłodowska-Curie Grant Agreement No. 823734, and is also co-funded by the European Union (project 101183150 - OCEANS).  We also thank Graeme Candlish and Omar Cuevas for allowing us to use part of their computer facilities for the initial calculation of our grid.

\end{acknowledgements}

%
%
\bibliographystyle{aa}
\bibliography{cites} 

\begin{appendix} 
\onecolumn
\section{Lines available in ISOSCELES}\label{appendixA}

The lines available in ISOSCELES are listed in Table \ref{optical} and Table \ref{ir} for the optical and infrared ranges, respectively.

\begin{table}[h!]
    \caption{List of optical spectral lines available in ISOSCELES}
    \label{optical}
    \centering
    \begin{tabular}{lllllllll}
        \hline
        \hline
        Ion & \multicolumn{8}{c}{$\lambda [\AA]$} \\
\hline
H\,I     & 6563 (H$\alpha$) & 4861 (H$\beta$) & 4340 (H$\gamma$) & 4102 (H$\delta$)  & 3970 (H$\epsilon$)  &  &  &  \\
He\,I    & 4026 & 4387 & 4471 & 4713 & 4922 & 6678 &  &  \\
He\,II   & 4200 & 4541 & 4686 & 6406 & 6527 & 6683 &  &  \\
Si\,II   & 4128 & 4130 & 5041 & 5056 &  &  &  &  \\
Si\,III  & 4552 & 4567 & 4574 & 4716 & 4813 & 4819 & 4829 & 5739 \\
Si\,IV   & 4089 & 4116 & 4212 & 4950 & 6667 & 6701 &  &  \\
        \hline
    \end{tabular}
\end{table}

\begin{table}[h!]
    \caption{List of infrared spectral lines available in ISOSCELES}
        \label{ir}
    \centering
\begin{tabular}{lllllllll}
\hline
\hline
Ion & \multicolumn{8}{c}{$\lambda [\AA]$} \\
\hline
He\,I    & 17000 & 20580 & 21100 &  &  &  &  &  \\
He\,II   & 11628 & 11677 & 15723 & 16922 & 21880 & 30908 &  &  \\
H\,I        & 18756 (P$\alpha$) & 12821 (P$\beta$) & 10941 (P$\gamma$) &  &  &  &  &  \\
H\,I       & 40522 (Br$\alpha$) & 26258 (Br$\beta$) & 21661 (Br$\gamma$) & 17362 (Br10) & 16811(Br11) & 16411 (Br12) &  &  \\
H\,I       & 37405 (Pf$\gamma$) & 32969 (Pf9 ) & 30392 (Pf10) &  &  &  &  &  \\    
        \hline
    \end{tabular}
\end{table}

\section{Uncertainties estimation sensitivity}
\label{uncertainties}

In this appendix, we present an analysis of the sensitivity of uncertainty estimates based on the number of top-ranked models included in the calculation. Models are sorted by increasing $\chi^2$ value, and uncertainties are derived from the distribution of a selected number $N$ of best-fitting models, following a quantile-based approach similar to that of \citet{Turis2025}. Specifically, we use HD\,206165 and HD\,14633 as test cases and compute the 0.25 quantile ($q_{0.25}$) for $N = 3{,}000$, $5{,}000$, $10{,}000$, $15{,}000$ and $20{,}000$. The results are presented in Table~\ref{tab:errors_hd}, additionally, the table includes the best-fit parameter values.

\begin{table*}[h!]
\caption{Uncertainty estimates as a function of the number of models $N$ with lowest $\chi^2$ values considered. The best-fit parameters for each star are listed in the second column for reference.}
\label{tab:errors_hd}
\centering
\begin{tabular}{lcccccc}
\hline\hline
HD 206165 & Best-fit & $N=3{,}000$ & $N=5{,}000$ & $N=10{,}000$ & $N=15{,}000$ & $N=20{,}000$ \\
\hline
$T_{\mathrm{eff}}$ [K]             & 20\,000 & 1294  & 1296  & 1319  & 1316  & 1326 \\
$\log g$ [dex]                     & 2.70    & 0.19  & 0.21  & 0.23  & 0.25  & 0.26 \\
$\dot{M}$ [$10^{-6} M_\odot$/yr]   & 0.175    & 2.11  & 5.55  & 12.61 & 13.99 & 15.22 \\
$v_\infty$ [km/s]                  & 398.55   & 41.1  & 50.8  & 73.6  & 82.5  & 87.6 \\
$\alpha$                           & 0.45    & 0.11  & 0.12  & 0.12  & 0.12  & 0.12 \\
$k$                                & 0.30    & 0.13  & 0.13  & 0.14  & 0.14  & 0.14 \\
$\delta$                           & 0.30    & 0.20  & 0.20  & 0.20  & 0.20  & 0.20 \\
$v_{\mathrm{mic}}$ [km/s]          & 10    & 2.70  & 3.10  & 4.09  & 4.75  & 5.15 \\
$\log \epsilon_{\rm{Si}}$          & 7.51    & 0.21  & 0.21  & 0.21  & 0.21  & 0.21 \\
\hline
HD 14633 & Best-fit & $N=3{,}000$ & $N=5{,}000$ & $N=10{,}000$ & $N=15{,}000$ & $N=20{,}000$ \\
\hline
$T_{\mathrm{eff}}$ [K]             & 36\,000 & 872   & 936   & 1016  & 1201  & 1424 \\
$\log g$ [dex]                     & 3.90    & 0.12  & 0.13  & 0.17  & 0.19  & 0.22 \\
$\dot{M}$ [$10^{-6} M_\odot$/yr]   & 0.12    & 1.88  & 3.01  & 7.72  & 8.69  & 9.15 \\
$v_\infty$ [km/s]                  & 1400.3    & 33.71 & 51.01 & 143.2 & 148.7 & 151.3 \\
$\alpha$                           & 0.61    & 0.13  & 0.12  & 0.12  & 0.12  & 0.12 \\
$k$                                & 0.10    & 0.11  & 0.11  & 0.13  & 0.14  & 0.14 \\
$\delta$                           & 0.20    & 0.11  & 0.10  & 0.10  & 0.10  & 0.10 \\
$v_{\mathrm{mic}}$ [km/s]          & 20      & 12.32 & 13.31 & 14.5  & 14.69  & 14.93 \\
$\log \epsilon_{\rm{Si}}$          & 7.51    & 0.20  & 0.21  & 0.21  & 0.21  & 0.21 \\
\hline
\end{tabular}
\end{table*}

As expected, the estimated uncertainties increase with $N$, especially for wind-related parameters such as $\dot{M}$ and $v_\infty$, which are more degenerate. However, the growth rate diminishes for large $N$, suggesting a saturation effect. For instance, the uncertainty in $\dot{M}$ for HD\,206165 increases significantly between $N=3{,}000$ and $N=10{,}000$, but shows a slower increment beyond this point. This behaviour implies that the quantile-based estimation (here $q_{0.25}$) stabilises as more of the relevant model space is sampled. 
Despite this stabilisation, we acknowledge that this method is not statistically rigorous and is sensitive to the grid's local density. Therefore, this analysis should be interpreted as an indicative rather than definitive error estimate. In future implementations, we plan to incorporate more robust uncertainty estimation techniques, such as Bayesian inference or profile likelihood approaches, to properly account for both model degeneracies and observational uncertainties.

\begin{figure}[h!]  
    \centering
\includegraphics[width=0.5\columnwidth]{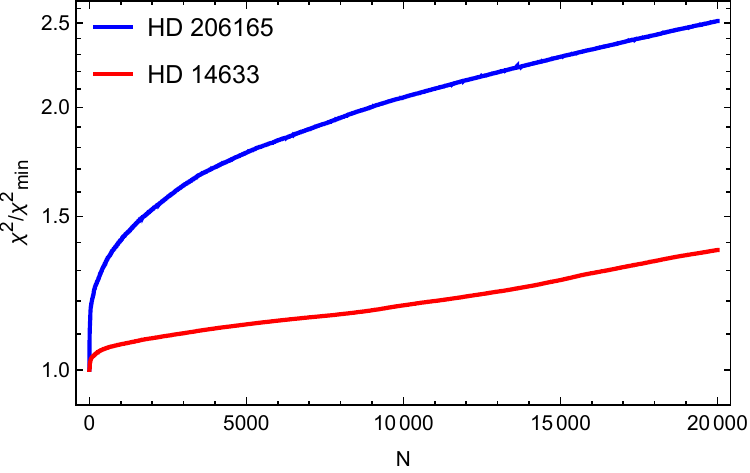}
        \caption{ Total $\chi^2$ value as a function of the number of selected models $N$ used to estimate uncertainties. As more models are included, the total $\chi^2$ increases steadily due to the incorporation of progressively less optimal fits.
    }
    \label{chi2N}   
\end{figure}

In addition to the numerical analysis, we include a figure (Fig.~\ref{chi2N}) showing the variation of the total $\chi^2$ value as a function of the number of selected models $N$. As expected, the $\chi^2$ value tends to increase with larger $N$, since more models with poorer fits are progressively included in the sample. This behaviour reinforces the idea that, while selecting a higher number of models may provide broader estimates of parameter uncertainties, it also introduces a growing contribution from suboptimal fits, which may bias the interpretation. Therefore, there is a compromise between uncertainty estimation and fitting quality, which must be considered when choosing the value of $N$.

To further investigate the diagnostic capabilities of individual spectral lines, we carried out an additional test using helium and silicon lines for the stars HD 206165  and HD 14633. In this controlled simulation, the wind-related parameters were held fixed at their best-fit values, and the effective temperature parameter was independently derived by fitting either only the helium lines or only the silicon lines.

This test enabled us to assess the constraining power of each group of lines on key parameters such as the effective temperature. For HD 206165, the silicon-only fit reproduced the same effective temperature as the full-line analysis ($20\,000$\,K), while the helium-only fit yielded a slightly lower value, differing by about 500\,K, which is within the estimated uncertainty. A similar outcome was obtained for HD 14633: the silicon-only fit recovered the global solution ($34\,000$\,K), and the helium-only fit returned a temperature of $33\,000$\,K, i.e. a difference of 1\,000\,K that also lies within the uncertainty range. These results confirm the robustness of the fitting procedure across different luminosity classes, from supergiants to dwarfs.

Although these results confirm the consistency of our method, implementing a weighted $\chi^2$ fitting approach that accounts for the varying sensitivity of spectral lines to specific parameters would require a broader set of diagnostics, including multiple ionisation stages. While such an extension lies beyond the scope of this study, it will be the focus of future work.

\section{Information about the observed stars}
\label{obs}
The information about the spectrographs, signal-to-noise (SNR), and observation dates is summarised in Table \ref{tobs}.

\begin{table}[h!]
\caption{Observation details for selected stars.}
\label{tobs}
    \centering
    \begin{tabular}{lccc}
        \hline
        \hline
        Star & Instrument & SNR  & Observation Date \\
        \hline
        HD 14633  & FIES   & 200  & 2009-11-13 \\
        HD 35299  & FEROS  & 303  & 2013-02-06\\
        HD 35912  & FIES   & 227  & 2008-11-08\\
        HD 14947  & FIES   & 130  & 2008-11-08\\
        HD 206165  & FIES  & 279  & 2009-11-10\\
        HD 53138  & FEROS  & 323  & 2005-04-22\\
        \hline
    \end{tabular}

    \label{tab:observations}
\end{table}

\pagebreak
\section{Spectral line fittings for the dwarf stars}
\label{dwarf-fig}

We present the comparisons between the observed spectra and the best-fitting ISOSCELES synthetic spectra for the dwarf stars in our sample. These figures illustrate the quality of the fits and complement the discussion in Section~\ref{section-dwarf}.

\begin{figure*}[h!]
\centering
\includegraphics[width=17cm]{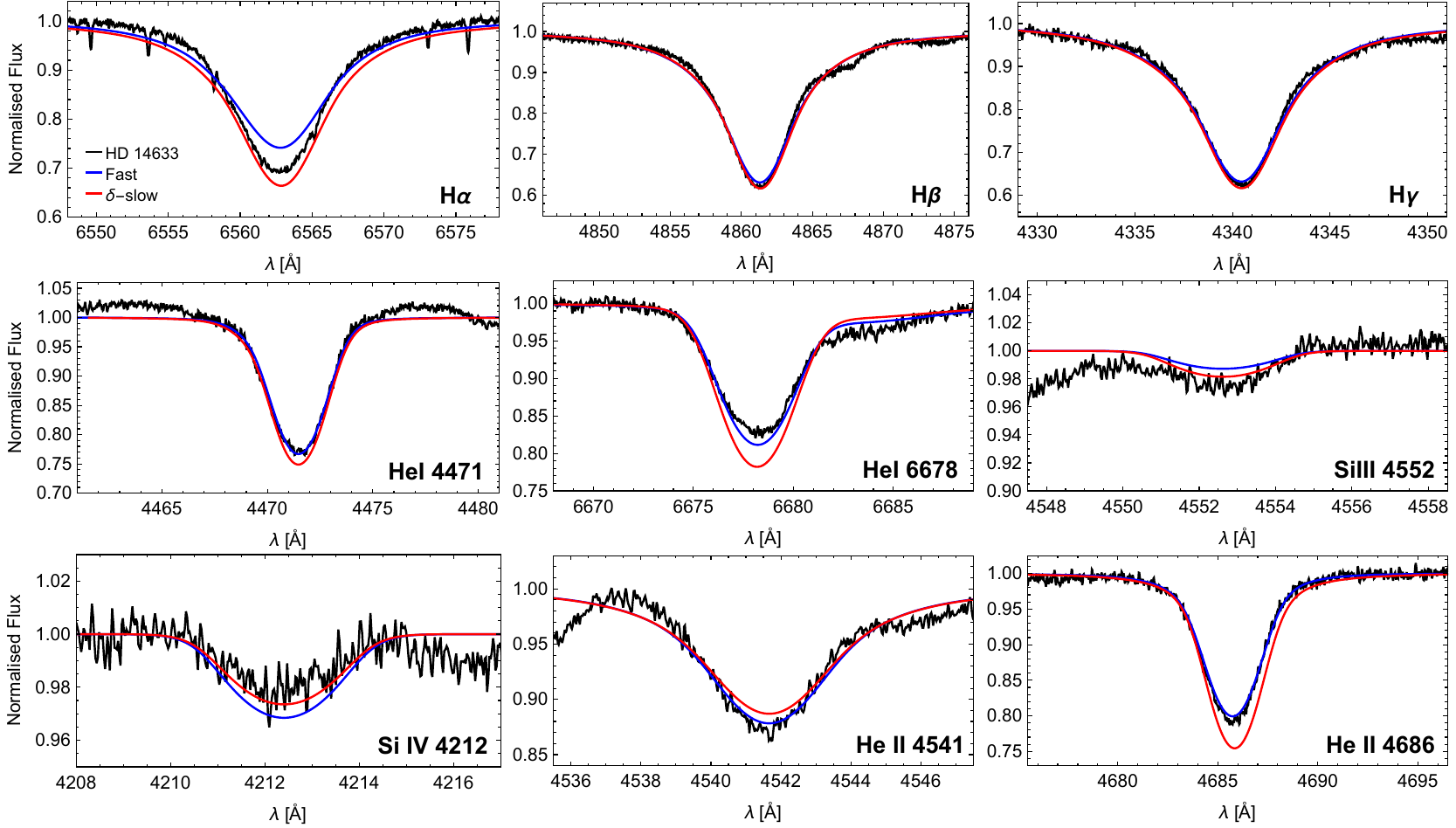}
        \caption{Observed line profiles of HD 14633 are shown in solid black. The best fast and $\delta$-slow solutions from ISOSCELES are plotted in blue and red, respectively.
 \label{HD14633}}
\end{figure*}

\begin{figure*}[h!]
\centering
\includegraphics[width=17cm]{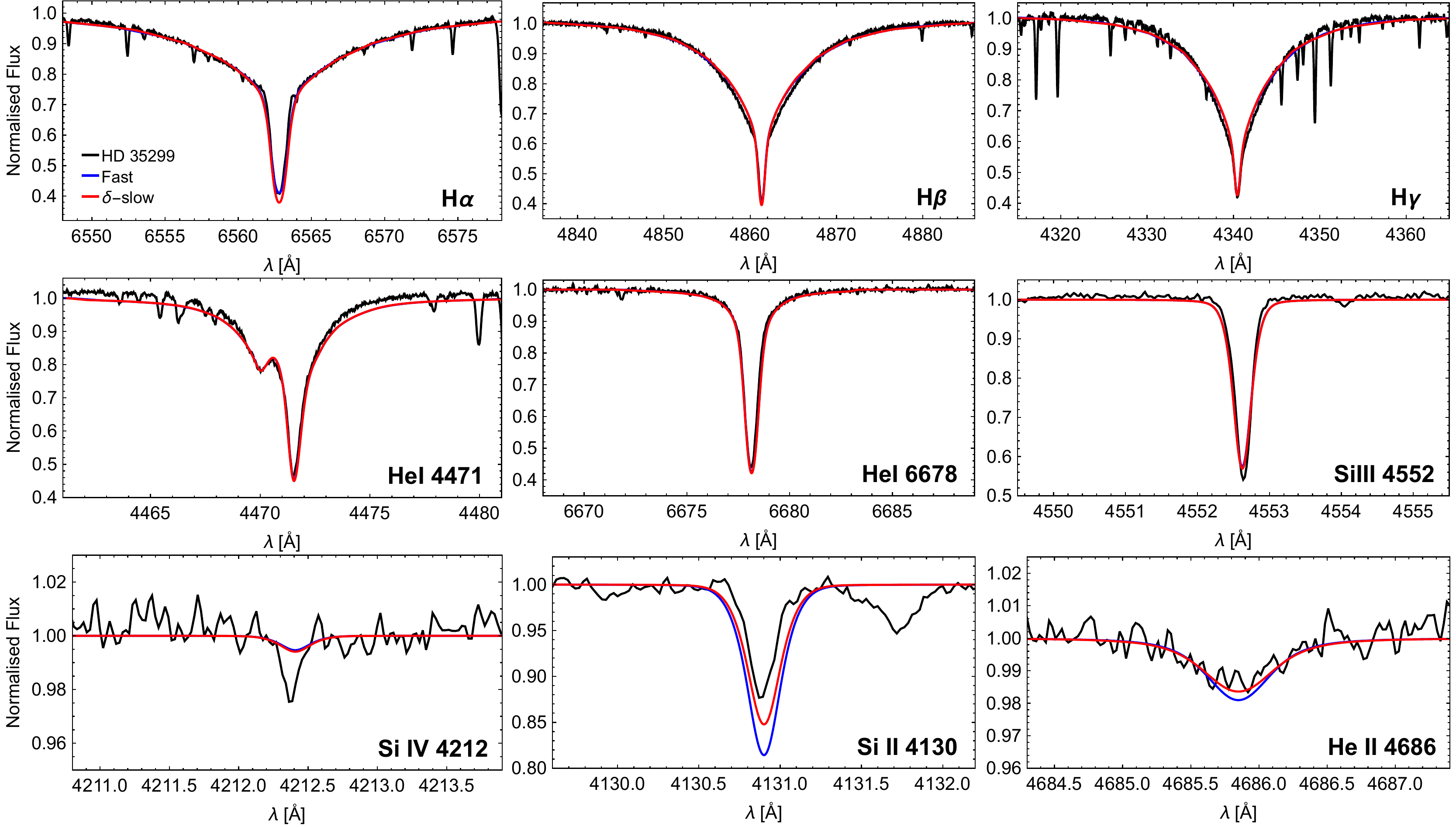}
        \caption{Same as Fig. \ref{HD14633}, but showing the observed and synthetic spectra for HD 35299.
 \label{HD35299}}
\end{figure*}

\begin{figure*}[h!]
\centering
\includegraphics[width=17cm]{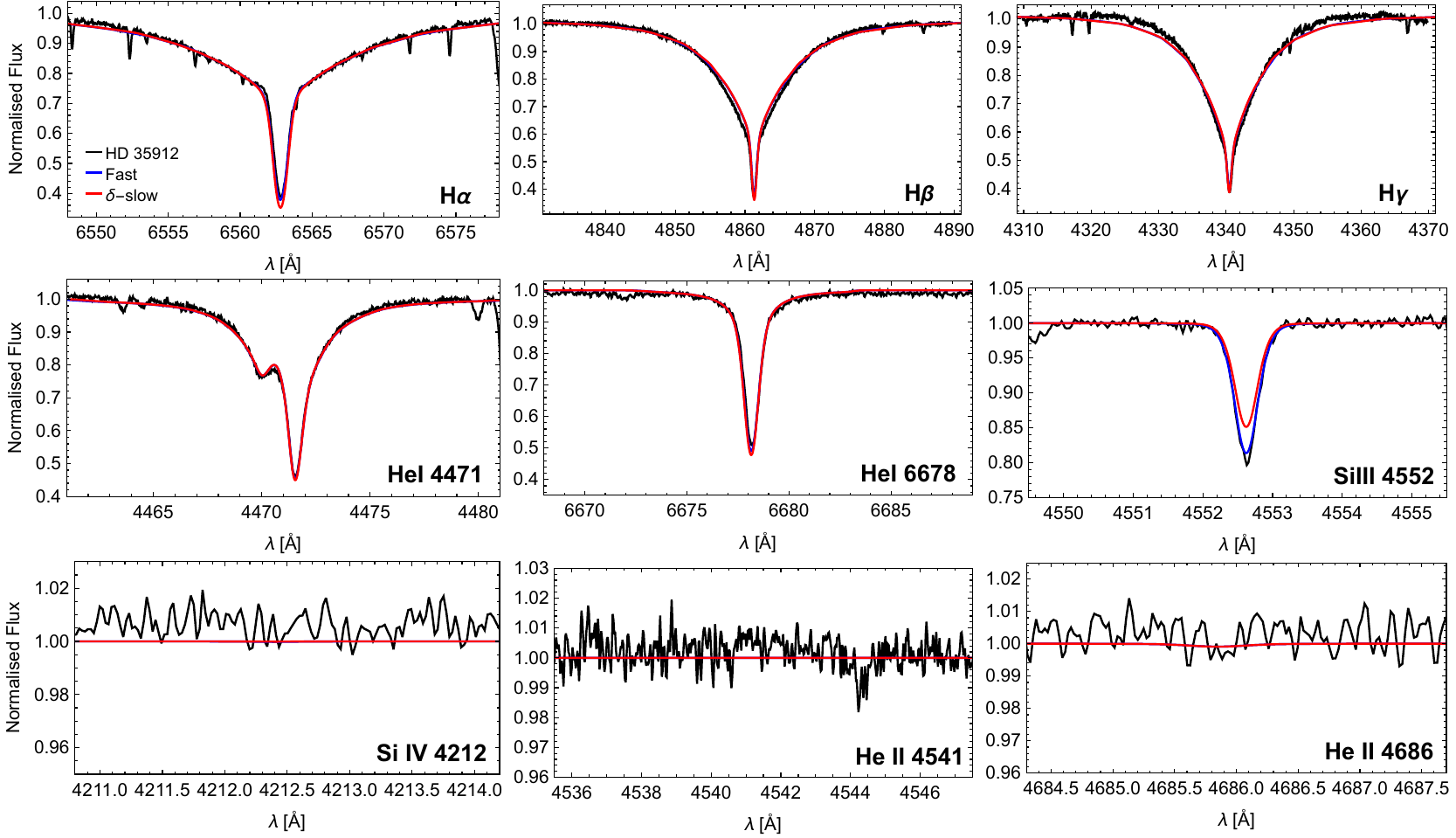}
        \caption{Same as Fig. \ref{HD14633}, but showing the observed and synthetic spectra for HD 35912.
 \label{HD35912}}
\end{figure*}

\end{appendix}

\end{document}